\documentclass[conference]{IEEEtran}
\IEEEoverridecommandlockouts
\usepackage{cite}
\usepackage{amsmath,amssymb,amsfonts}
\usepackage{algorithmic}
\usepackage{graphicx}
\usepackage{textcomp}
\usepackage{xcolor}
\usepackage{multirow}
\def\BibTeX{{\rm B\kern-.05em{\sc i\kern-.025em b}\kern-.08em
    T\kern-.1667em\lower.7ex\hbox{E}\kern-.125emX}}
\begin{document}

\title{Subjective evaluation of UHD video coded using VVC with LCEVC and ML-VVC}

\author{
    \IEEEauthorblockN{Naeem Ramzan, SMIEEE, Muhammad Tufail Khan
\\School of Computing, Engineering and Physical Sciences \\ University of the West of Scotland \\}
    \IEEEauthorblockA{naeem.ramzan@uws.ac.uk}
}

\maketitle

\begin{abstract}
This paper presents the results of a subjective quality assessment of a multilayer video coding configuration in which Low Complexity Enhancement Video Coding (LCEVC) is applied as an enhancement layer on top of a Versatile Video Coding (VVC) base layer. The evaluation follows the same test methodology and conditions previously defined for MPEG multilayer video coding assessments, with the LCEVC enhancement layer encoded using version 8.1 of the LCEVC Test Model (LTM).

The test compares reconstructed UHD output generated from an HD VVC base layer with LCEVC enhancement against two reference cases: upsampled VVC base layer decoding and multilayer VVC (ML-VVC). Two operating points are considered, corresponding to enhancement layers representing approximately 10\% and 50\% of the total bitrate. Subjective assessment was conducted using the Degradation Category Rating (DCR) methodology with twenty five participants, across a dataset comprising fifteen SDR and HDR sequences.

The reported results include Mean Opinion Scores (MOS) with associated 95\% confidence intervals, enabling comparison of perceptual quality across coding approaches and operating points within the defined test scope.
\end{abstract}


\section{Introduction}
In \cite{bn6} the results for the subjective testing according to the multilayer video coding test plan \cite{b6} were reported. The test exercised three types of MPEG multilayer video coding technologies which have different characteristics in terms of features, performance, and complexity, and are capable of dual-layer coding with an HD base layer and a UHD enhancement layer (2x spatial scalability). These comprise SHVC (the scalable extension of HEVC), the VVC multilayer profile, and LCEVC.
In that test, LCEVC was tested using a version of the reference software which was released in January 2025, namely LTM 7.0 \cite{b7}, using some constrained configurations.
Since the publication of the report, a new version of the LTM has been released, LTM 8.1 \cite{b9}. In this paper, we provide the results of a test done with the exact same methodology used in \cite{bn6} and \cite{b6} but using LTM 8.1 with the configuration as specified in the reference software manual.

\section{Methodology}
This section describes the methodology used in the tests reported here. 

\subsection{Dataset}
Table~\ref{tab1} provides the list of sequences used in this paper. More details about those sequences can be found in \cite{b6}.  

\begin{table}[htbp]
\caption{List of sequences}
\begin{center}
\renewcommand{\arraystretch}{1.2}
\begin{tabular}{|l|c|c|c|}
\hline
\textbf{Sequence} & \textbf{Resolution} & \textbf{Frames and fps} & \textbf{Colour Space} \\
\hline
BodeMuseum           & 3840×2160 & 600 (60 fps) & SDR, BT.2020 \\
Metro                & 3840×2160 & 600 (60 fps) & SDR, BT.2020 \\
OberbaumSpree        & 3840×2160 & 600 (60 fps) & SDR, BT.2020 \\
SubwayTree           & 3840×2160 & 600 (60 fps) & SDR, BT.2020 \\
WaterFront           & 3840×2160 & 240 (24 fps) & SDR, BT.2020 \\
FootballLargeAdvert  & 3840×2160 & 600 (60 fps) & HLG, BT.2020 \\
AMS01             & 3840×2160 & 600 (60 fps) & HLG, BT.2020 \\
AMS02             & 3840×2160 & 600 (60 fps) & HLG, BT.2020 \\
AMS05             & 3840×2160 & 600 (60 fps) & HLG, BT.2020 \\
WomenFootball        & 3840×2160 & 600 (60 fps) & PQ, BT.2020 \\
GreenMountains1      & 3840×2160 & 250 (25 fps) & PQ, BT.2020 \\
KitchenPressin       & 3840×2160 & 600 (60 fps) & PQ, BT.2020 \\
RiverPlate1          & 3840×2160 & 600 (60 fps) & PQ, BT.2020 \\
TiergartenParkway    & 3840×2160 & 600 (60 fps) & PQ, BT.2020 \\
WalkInPark           & 3840×2160 & 600 (60 fps) & PQ, BT.2020 \\
\hline
\end{tabular}
\label{tab1}
\end{center}
\end{table}

\subsection{Brief decription of the test}
The test was conducted to evaluate LCEVC combined with a VVC base layer against the upsampled VVC base layer. The resolution of the tested content is HD (VVC base layer) + UHD (LCEVC enhancement layer), with the HD to UHS upsampled version of the VVC base layer. 
Each enhancement layer is added on top of the base layer to achieve one of two target operating points: (a) E10: Enhancement layer represents 10\% of the total bitrate; (b) E50: Enhancement layer represents 50\% of the total bitrate.
In order to provide calibration points with respect to the tests reported in \cite{bn6}, the test also includes an evaluation of the upsampled base and ML-VVC.

\subsection{Configurations}
The VVC base layer and ML-VVC enhancement layer in this test was encoded using VTM 23.8 \cite{b8} with the CTC RA configuration and corresponding SDR \cite{bn3} and HDR \cite{bn4} CTC testing conditions. 
In this test, a setting with flat quantization matrices and disabled local quantization adaptation was used. The configurations used for this test are the same as those used in \cite{b6}.
The LCEVC enhancement layer in this test was encoded using LTM 8.1 with the configurations as described in \cite{b9}. 
The downsampler and upsampler used for generating the base layer and upsampling it correspond to those described in \cite{b6}.

\begin{table}[t]
\centering
\caption{Quantization parameters}
\label{tab:quant_lcevc_only}
\renewcommand{\arraystretch}{1.2}
\begin{tabular}{|l|c|c|c|c|c|}
\hline
\multirow{2}{*}{Sequence name} &
\multirow{2}{*}{Enh \%} &
\multirow{2}{*}{\shortstack{BL\\QP}}&
\multirow{2}{*}{\shortstack{ML-VVC\\QP}} &
\multicolumn{2}{|c|}{LCEVC} \\
\cline{5-6}
& & & & SW 2 & SW 1 \\
\hline

\multirow{2}{*}{BodeMuseum} & 10\% & \multirow{2}{*}{29} & 41 & 4100 & 1000 \\
                           & 50\% &                     & 34 & 1250 & 3000 \\
\hline
\multirow{2}{*}{Metro} & 10\% & \multirow{2}{*}{29} & 39 & 2750 & 32767 \\
                       & 50\% &                     & 33 & 1000 & 32767 \\
\hline
\multirow{2}{*}{OberbaumSpree} & 10\% & \multirow{2}{*}{29} & 37 & 2500 & 32767 \\
                              & 50\% &                     & 33 &  900 & 1000 \\
\hline
\multirow{2}{*}{SubWayTree} & 10\% & \multirow{2}{*}{31} & 41 & 3000 & 32767 \\
                            & 50\% &                     & 35 & 1500 & 32767 \\
\hline
\multirow{2}{*}{WaterFront} & 10\% & \multirow{2}{*}{27} & 43 & 3500 & 32767 \\
                            & 50\% &                     & 35 & 1500 & 32767 \\
\hline
\multirow{2}{*}{FootballLargeAdvert} & 10\% & \multirow{2}{*}{31} & 40 & 2750 & 32767 \\
                                     & 50\% &                     & 34 & 1100 & 32767 \\
\hline
\multirow{2}{*}{H3\_AMS01} & 10\% & \multirow{2}{*}{29} & 42 & 3000 & 32767 \\
                           & 50\% &                     & 35 & 1400 & 32767 \\
\hline
\multirow{2}{*}{H3\_AMS02} & 10\% & \multirow{2}{*}{29} & 41 & 3000 & 32767 \\
                           & 50\% &                     & 35 & 1300 & 32767 \\
\hline
\multirow{2}{*}{H3\_AMS05} & 10\% & \multirow{2}{*}{29} & 41 & 2500 & 32767 \\
                           & 50\% &                     & 34 & 1250 & 1250 \\
\hline
\multirow{2}{*}{WomenFootball} & 10\% & \multirow{2}{*}{31} & 38 & 1750 & 32767 \\
                               & 50\% &                     & 34 &  700 & 32767 \\
\hline
\multirow{2}{*}{GreenMountain1} & 10\% & \multirow{2}{*}{27} & 35 & 3000 & 32767 \\
                                & 50\% &                     & 31 & 1000 & 32767 \\
\hline
\multirow{2}{*}{KitchenDressing} & 10\% & \multirow{2}{*}{27} & 39 & 2500 & 3000 \\
                                 & 50\% &                     & 30 &  600 & 32767 \\
\hline
\multirow{2}{*}{RiverPlate\_1} & 10\% & \multirow{2}{*}{27} & 38 & 2250 & 32767 \\
                               & 50\% &                     & 31 &  900 & 32767 \\
\hline
\multirow{2}{*}{TiergartenParkway} & 10\% & \multirow{2}{*}{29} & 38 & 3000 & 32767 \\
                                   & 50\% &                     & 33 & 1000 & 32767 \\
\hline
\multirow{2}{*}{WalkInThePark} & 10\% & \multirow{2}{*}{27} & 34 & 2500 & 3000 \\
                               & 50\% &                     & 30 &  750 & 32767 \\
\hline
\end{tabular}
\vspace{0.5em}
\end{table}

\subsection{Quantization parameters}
The quantization parameters used in this test are reported in Table ~\ref{tab:quant_lcevc_only}. In particular, the Quantization Parameters (QPs) used for the base layer and ML-VVC correspond to those used in \cite{b6}. For LCEVC, instead, the two standard quantization parameters, SW1 and SW2, were defined using a methodology based on a convex hull method to determine the best pair of quantization parameters given a specific rate and distortion. 

\subsection{Test design and execution}
The test was conducted in the viewing lab of the University of the West of Scotland. 
The evaluation followed the Degradation Category Rating (DCR) methodology. The test was conducted blindly, with anonymized encodings and in randomized order, involving 25 subjects. The assessment approach closely follows the methodology described in \cite{b6}.

Participants rated visual quality using an 11-point scale ranging from 0 (lowest quality) to 10 (highest quality) as shown in Table ~\ref{tab11grade}.

\begin{table}[htbp]
\centering
\caption{11-grade impairment scale}
\renewcommand{\arraystretch}{1.2}
\begin{tabular}{|c|l|l|}
\hline
\textbf{Score} & \multicolumn{2}{c|}{\textbf{Impairment item}} \\ \hline

10 & Imperceptible & \\ \hline

9 & \multirow{2}{*}{Slightly perceptible} & somewhere \\ \cline{1-1}\cline{3-3}
8 &  & everywhere \\ \hline

7 & \multirow{2}{*}{Perceptible} & somewhere \\ \cline{1-1}\cline{3-3}
6 &  & everywhere \\ \hline

5 & \multirow{2}{*}{Clearly perceptible} & somewhere \\ \cline{1-1}\cline{3-3}
4 &  & everywhere \\ \hline

3 & \multirow{2}{*}{Annoying} & somewhere \\ \cline{1-1}\cline{3-3}
2 &  & everywhere \\ \hline

1 & \multirow{2}{*}{Severely annoying} & somewhere \\ \cline{1-1}\cline{3-3}
0 &  & everywhere \\ \hline
\end{tabular}
\label{tab11grade}
\end{table}

\section{Experimental Results}

\begin{figure}[htbp]
\centerline
{\includegraphics[width=0.50\textwidth]{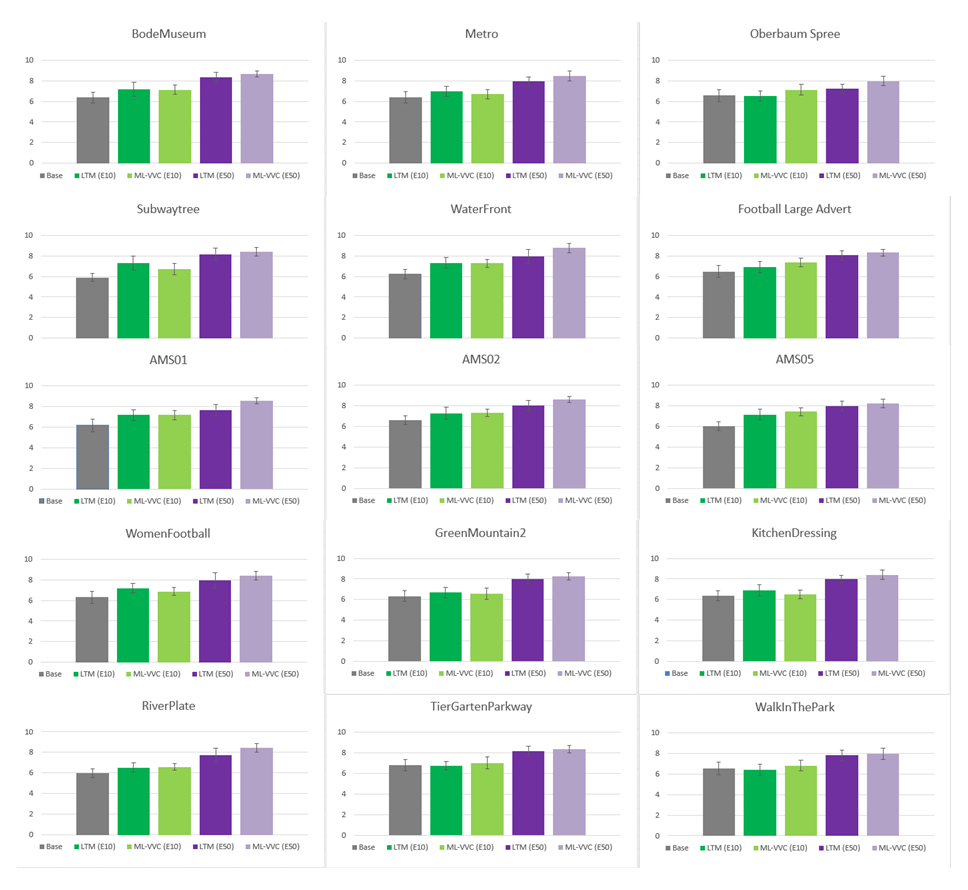}}  
\caption{Plots reporting, for each sequence, the MOS and CI for the three schemes: Base (grey), LCEVC E10 (dark green), ML-VVC E10 (light green), LCEVC E50 (dark purple) and ML-VVC E50 (light purple)}
\label{fig4}
\end{figure}

Figure~\ref{fig4} shows the plots corresponsing to the experimental results. In addition, Table ~\ref{tab4} and Table ~\ref{tab5} report the resulting overall MOS scores with the 95\% confidence interval (CI) marked for both, the positive and negative direction. The results are shown in a grouped manner for the cases of the enhancement layer (EL) 10\% and 50\% (E10 / E50). 

\begin{table}[htbp]
\caption{Comparison between multilayer approaches (E10)}
\begin{center}
\renewcommand{\arraystretch}{1.2}
\begin{tabular}{|l|cc|cc|cc|}
\hline
\textbf{Sequence Name} & \multicolumn{2}{c|}{\textbf{Base}} & \multicolumn{2}{c|}{\textbf{LCEVC E10}} & \multicolumn{2}{c|}{\textbf{ML-VVC E10}} \\
 & MOS & CI & MOS & CI & MOS & CI \\
\hline
Bode Museum & 6.36 & 0.51 & 7.16 & 0.65 & 7.12 & 0.46 \\
Metro & 6.40 & 0.57 & 6.96 & 0.50 & 6.68 & 0.44 \\
OberbaumSpree & 6.56 & 0.59 & 6.52 & 0.42 & 7.12 & 0.52 \\
Subwaytree & 5.88 & 0.39 & 7.28 & 0.69 & 6.72 & 0.53 \\
WaterFront & 6.24 & 0.46 & 7.32 & 0.50 & 7.28 & 0.40 \\
FootBallLargeAdvert & 6.48 & 0.57 & 6.88 & 0.55 & 7.36 & 0.45 \\
AMS01 & 6.16 & 0.61 & 7.16 & 0.50 & 7.16 & 0.44 \\
AMS02 & 6.64 & 0.41 & 7.28 & 0.58 & 7.32 & 0.37 \\
AMS05 & 6.04 & 0.40 & 7.16 & 0.53 & 7.44 & 0.38 \\
WomenFootball & 6.32 & 0.58 & 7.20 & 0.45 & 6.88 & 0.39 \\
GreenMountains1 & 6.32 & 0.53 & 6.68 & 0.53 & 6.56 & 0.57 \\
KitchenDressing & 6.40 & 0.48 & 6.88 & 0.55 & 6.52 & 0.42 \\
RiverPlate & 6.00 & 0.42 & 6.52 & 0.46 & 6.60 & 0.34 \\
TierGartenParkway & 6.80 & 0.54 & 6.76 & 0.43 & 7.04 & 0.60 \\
WalkInThePark & 6.52 & 0.59 & 6.40 & 0.57 & 6.80 & 0.51 \\
\hline
\end{tabular}
\label{tab4}
\end{center}
\end{table}

\begin{table}[htbp]
\caption{Comparison between multilayer approaches (E50)}
\begin{center}
\renewcommand{\arraystretch}{1.2}
\begin{tabular}{|l|cc|cc|cc|}
\hline
\textbf{Sequence Name} & \multicolumn{2}{c|}{\textbf{Base}} & \multicolumn{2}{c|}{\textbf{LCEVC E50}} & \multicolumn{2}{c|}{\textbf{ML-VVC E50}} \\
 & MOS & CI & MOS & CI & MOS & CI \\
\hline
Bode Museum & 6.36 & 0.51 & 8.32 & 0.47 & 8.64 & 0.29 \\
Metro & 6.40 & 0.57 & 7.92 & 0.45 & 8.48 & 0.49 \\
OberbaumSpree & 6.56 & 0.59 & 7.20 & 0.48 & 7.96 & 0.43 \\
Subwaytree & 5.88 & 0.39 & 8.16 & 0.58 & 8.40 & 0.42 \\
WaterFront & 6.24 & 0.46 & 7.96 & 0.67 & 8.76 & 0.46 \\
FootBallLargeAdvert & 6.48 & 0.57 & 8.04 & 0.47 & 8.32 & 0.33 \\
AMS01 & 6.16 & 0.61 & 7.64 & 0.54 & 8.52 & 0.30 \\
AMS02 & 6.64 & 0.41 & 8.04 & 0.47 & 8.64 & 0.29 \\
AMS05 & 6.04 & 0.40 & 8.00 & 0.48 & 8.24 & 0.43 \\
WomenFootball & 6.32 & 0.58 & 7.96 & 0.73 & 8.44 & 0.42 \\
GreenMountains1 & 6.32 & 0.53 & 8.00 & 0.48 & 8.24 & 0.35 \\
KitchenDressing & 6.40 & 0.48 & 8.00 & 0.38 & 8.40 & 0.45 \\
RiverPlate & 6.00 & 0.42 & 7.76 & 0.67 & 8.44 & 0.42 \\
TierGartenParkway & 6.80 & 0.54 & 8.20 & 0.45 & 8.36 & 0.38 \\
WalkInThePark & 6.52 & 0.59 & 7.80 & 0.54 & 7.96 & 0.53 \\
\hline
\end{tabular}
\label{tab5}
\end{center}
\end{table}

\section{Conclusion}
This study re-examined the subjective performance of an LCEVC enhancement layer applied to a VVC base layer, using LTM 8.1 and the same test plan and laboratory conditions previously employed for MPEG multilayer video coding evaluations. The assessment covered a set of fifteen UHD SDR and HDR sequences and considered two enhancement bitrate operating points, E10 and E50, relative to the total bitrate.

Across the tested content, both LCEVC-based enhancement and ML-VVC consistently achieved higher subjective quality scores than simple upsampling of the VVC base layer. At the E10 operating point, the observed differences in MOS between enhancement approaches and the upsampled base layer were generally positive, although sequence-dependent variations were observed and, in some cases, differences remained within the reported 95\% confidence intervals. At the E50 operating point, higher MOS values were observed for both multilayer approaches across all sequences, reflecting the increased bitrate allocated to enhancement.

When comparing LCEVC and ML-VVC directly, the results indicate that both approaches deliver comparable subjective quality at equivalent operating points within the limits of the test design. Differences between the two enhancement methods were sequence-dependent and, in most cases, fell within the associated confidence intervals, indicating no statistically robust separation under the tested conditions.

It should be noted that these findings are specific to the selected configurations, encoder versions, quantization settings, and down-/up-sampling processes defined by the test plan. The results do not address encoder or decoder complexity, implementation cost, latency, or performance outside the evaluated operating points. Further investigations could extend this analysis to additional bitrate points, alternative resampling methods, and different encoder configurations in order to more comprehensively characterize trade-offs between multilayer coding approaches.

\section{Acknowledgments}
We acknowledge the V-Nova team for providing the encoder configurations, quantization parameter settings, and the encoded bitstreams used in this evaluation. All materials were supplied in an anonymised form, together with scripts that enabled the laboratory to decode the bitstreams automatically without prior knowledge of the technologies associated with each stream.

\vspace{12pt}

\end{document}